\makeatletter
\def\input@path{{Definitions/}}
\makeatother
\documentclass[aps,pra,reprint,groupedaddress,showkeys,amsmath,amssymb,floatfix,nobibnotes]{revtex4-2}
\usepackage[T1]{fontenc}
\usepackage[utf8]{inputenc}
\usepackage{graphicx}
\usepackage{bm}
\usepackage{makecell}
\usepackage{comment}
\usepackage{soul}
\usepackage{xcolor}
\usepackage{hyperref}
\hypersetup{colorlinks=true,linkcolor=blue,citecolor=blue,urlcolor=blue}
\makeatletter
\let\auto@bib@innerbib\@empty
\let\write@bibliographystyle\relax
\makeatother
\newcommand{\authorcontributions}[1]{\par\smallskip\noindent\textbf{Author Contributions:} #1\par}
\newcommand{\funding}[1]{\par\smallskip\noindent\textbf{Funding:} #1\par}
\newcommand{\institutionalreview}[1]{\par\smallskip\noindent\textbf{Institutional Review Board Statement:} #1\par}
\newcommand{\informedconsent}[1]{\par\smallskip\noindent\textbf{Informed Consent Statement:} #1\par}
\newcommand{\dataavailability}[1]{\par\smallskip\noindent\textbf{Data Availability Statement:} #1\par}
\newcommand{\conflictsofinterest}[1]{\par\smallskip\noindent\textbf{Conflicts of Interest:} #1\par}
\begin{document}
\title{Size-Dependent High-Order Harmonic Generation from Linear Carbon-Chain Molecules in Linearly Polarized Laser Fields}
\author{Yifan Kang}
\author{Yang Yang}
\author{Yingtan Li}
\author{Shuwen Pang}
\author{Huiwen Yang}
\author{Yingdan Zhang}
\author{Shushan Zhou}
\email[Correspondence: ]{zhoushushan@lnnu.edu.cn}
\affiliation{School of Physics and Electronic Technology, Liaoning Normal University, Dalian 116029, China}
\begin{abstract}
High-order harmonic generation from molecules is highly sensitive to molecular geometry and strong-field electron dynamics, providing a powerful route for probing and controlling ultrafast processes in extended systems. In this work, we systematically investigate the influence of molecular chain length on HHG from linear carbon-chain molecules C$_2$H$_2$, C$_6$H$_2$, and C$_{10}$H$_2$ driven by an 800 nm linearly polarized laser field. To isolate the molecular-size effect from the influence of different ionization potentials, the laser intensities are adjusted to maintain the same Keldysh parameter for different molecules. Real-time time-dependent density functional theory simulations are performed with the molecular axis aligned parallel to the laser polarization direction. The results show that increasing the molecular chain length significantly enhances the harmonic yield and tends to extend the cutoff region. In particular, C$_{10}$H$_2$ exhibits much stronger plateau harmonics than C$_2$H$_2$, while the intermediate molecule C$_6$H$_2$ shows harmonic intensities between those of C$_2$H$_2$ and C$_{10}$H$_2$. Time-frequency analysis and charge-density evolution reveal that the enhanced HHG emission in longer chains originates from the enlarged electron acceleration distance and increased probability of multi-center recollision, especially near the terminal carbon atoms. These findings demonstrate that molecular-size engineering offers an effective strategy for controlling harmonic emission and optimizing attosecond light sources in extended molecular systems.

\end{abstract}
\maketitle

\section{Introduction}

High-order harmonic generation (HHG), driven by the interaction between intense laser fields and matter, provides a fundamental route to coherent extreme-ultraviolet radiation and attosecond pulse generation \cite{1,2,3,4,5}. In the semiclassical three-step model, an electron is first ionized by the strong laser field, then accelerated in the continuum, and finally driven back to recombine with its parent ion, emitting high-energy photons \cite{6,7,8,9,10,11,12,13,14,15}. For atomic systems, this model predicts the well-known cutoff law, $E_{\mathrm{cutoff}} = I_p + 3.17U_p$,
where $I_p$ is the ionization potential and $U_p$ is the ponderomotive energy \cite{16}. This cutoff law has been widely verified in atoms and serves as the basic theoretical framework for understanding HHG.

When the target is extended from atoms to molecules, HHG becomes sensitive to molecular structure and electron dynamics. The multi-center geometry, orbital symmetry, and nuclear degrees of freedom allow molecular harmonic spectra to encode information on molecular orbitals, ultrafast electron motion, and electron--nuclear coupling processes \cite{16}. In molecules, the ionized electron may recombine not only with the parent ion but also with neighboring atomic centers, leading to multi-center interference and structure-dependent spectral features \cite{17}. These characteristics make molecular HHG a powerful tool for probing molecular structure and ultrafast dynamics.

Extensive studies have been carried out on HHG from diatomic molecules \cite{18,19,20,21,22} and simple polyatomic molecules \cite{23,24,25,26,27}. These works have revealed the important roles of molecular alignment, multi-orbital contributions, and tailored laser fields in controlling harmonic emission. For example, molecular orientation can strongly influence electron recollision pathways \cite{21}, higher occupied orbitals may contribute to cutoff extension under suitable laser conditions \cite{25}, and synthesized two-color fields can manipulate molecular electron wave-packet dynamics \cite{27}. Nevertheless, for spatially extended molecules, especially one-dimensional chain-like systems, the influence of molecular size on harmonic intensity and cutoff behavior remains insufficiently understood.
Linear carbon-chain molecules provide an ideal platform for investigating size-dependent HHG. Their quasi-one-dimensional geometry, tunable chain length, and delocalized $\pi$-electron structure can facilitate electron acceleration and recollision along the molecular axis. Compared with small molecules, longer chains may provide an extended acceleration region and more recollision centers, thereby modifying electron trajectories, enhancing harmonic emission, and possibly shifting the cutoff to higher photon energies \cite{33,34}. However, changes in molecular size are usually accompanied by variations in ionization potential. Therefore, it is necessary to separate pure geometric effects from ionization effects. The Keldysh parameter, $\gamma = \sqrt{\frac{I_p}{2U_p}}$, provides a useful criterion for characterizing the ionization regime. By adjusting the laser intensity to keep $\gamma$ identical for different molecules, one can compare their HHG spectra under comparable ionization conditions \cite{32}.

In this work, we systematically investigate HHG from linear carbon-chain molecules with different lengths, namely C$_2$H$_2$, C$_6$H$_2$, and C$_{10}$H$_2$, driven by an 800 nm linearly polarized laser field. The molecular axis is aligned parallel to the laser polarization direction, and the laser intensity is chosen to maintain the same Keldysh parameter for different molecules. Time-dependent density functional theory (TDDFT) is employed to simulate the strong-field electron dynamics and harmonic emission \cite{35,36,37,38}. By comparing the harmonic spectra, time-frequency distributions, and charge-density evolution, we clarify how molecular chain length affects harmonic yield and cutoff behavior. Our results show that increasing molecular size can significantly enhance the harmonic intensity and promote cutoff extension, mainly due to the enlarged electron acceleration distance and enhanced multi-center recollision probability.

\section{Results and Discussion}\label{sec2}


\begin{figure}[t]
\centering
\includegraphics[width=0.94\columnwidth]{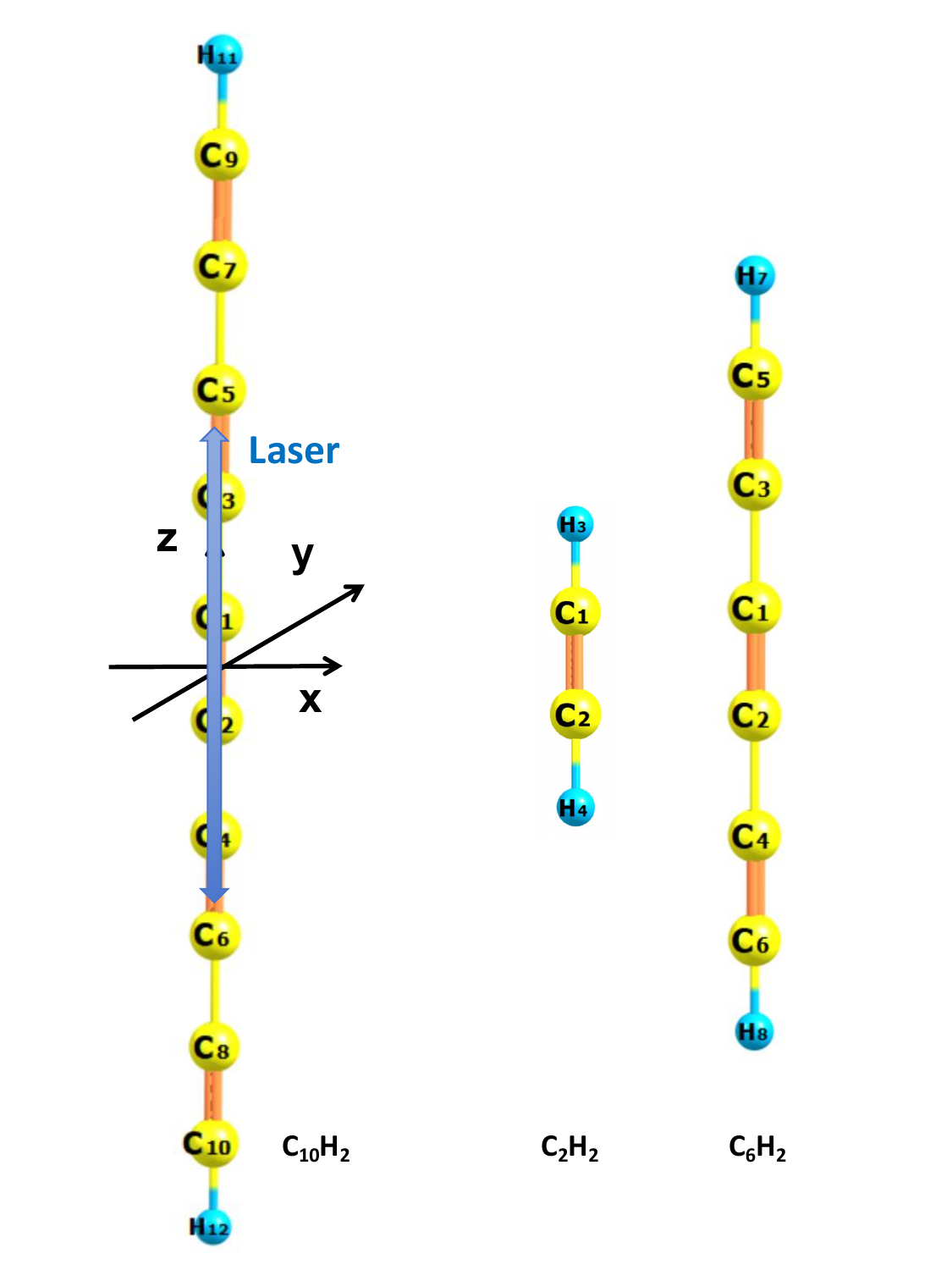}
\caption{
Schematic illustration of the molecular orientation with respect to the linearly polarized laser field. The molecular axes of C$_2$H$_2$, C$_6$H$_2$, and C$_{10}$H$_2$ are aligned along the $z$ axis, parallel to the laser polarization direction. The molecular lengths are 3.15 a.u., 8.01 a.u., and 12.87 a.u., respectively.
}
\label{fig:1}
\end{figure}


We first compare the HHG spectra of C$_2$H$_2$ and C$_{10}$H$_2$ under the same Keldysh parameter to clarify the influence of molecular size. For C$_{10}$H$_2$, the laser-field amplitude is chosen as 0.057 a.u., corresponding to $\gamma=0.724$. Using the same value of $\gamma$, the equivalent field amplitude for C$_2$H$_2$ is determined to be 0.073 a.u. The laser wavelength is fixed at 800 nm, and the molecular axis is parallel to the laser polarization direction.

Figure~2 shows the calculated HHG spectra of C$_2$H$_2$ and C$_{10}$H$_2$. Although the two molecules are compared under equivalent ionization conditions, their harmonic spectra exhibit clear differences. The longer molecule C$_{10}$H$_2$ produces a significantly stronger harmonic signal in the plateau region than C$_2$H$_2$. In addition, the cutoff region of C$_{10}$H$_2$ tends to extend toward higher harmonic orders, indicating that the enlarged molecular size modifies the electron acceleration and recollision dynamics.

The cutoff position in the atomic three-step picture can be estimated by $N_{\mathrm{cutoff}} = \frac{I_p+3.17U_p}{\omega_0}$, where $U_p = \frac{E_0^2}{4\omega_0^2}$ is the ponderomotive energy. However, for extended molecular systems, the actual cutoff behavior may deviate from this simple prediction because molecular HHG involves multi-center potentials, orbital delocalization, and possible contributions from multiple occupied orbitals. Therefore, the intensity enhancement and cutoff variation observed here should be understood as the combined result of laser-driven electron acceleration in an extended molecular potential and multi-center recollision.

\begin{figure}[t]
\centering
\includegraphics[width=0.94\columnwidth]{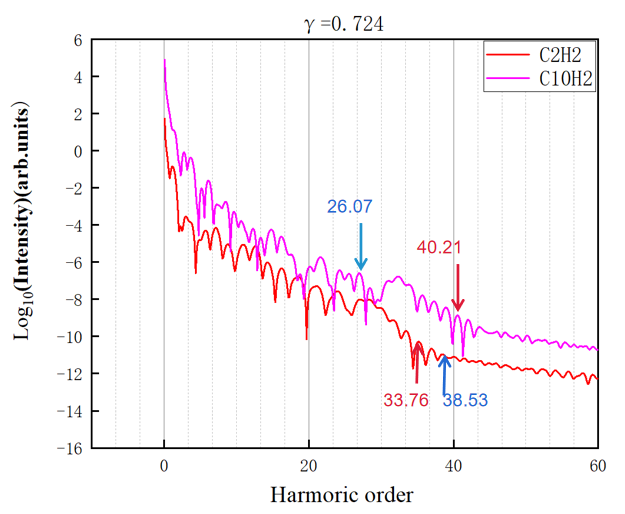}
\caption{
HHG spectra of C$_2$H$_2$ and C$_{10}$H$_2$ driven by an 800 nm linearly polarized laser field. The molecular axis is parallel to the laser polarization direction, as shown in Fig.~1. The electric-field amplitude is 0.057 a.u. for C$_{10}$H$_2$ and 0.073 a.u. for C$_2$H$_2$, corresponding to the same Keldysh parameter $\gamma=0.724$. The laser pulse has a trapezoidal envelope with a one-cycle ramp-on, a four-cycle plateau, and a one-cycle ramp-off.
}
\label{fig:2}
\end{figure}


To understand the microscopic origin of the enhanced harmonic emission in the longer molecule, we further analyze the time-frequency distribution and the time-dependent charge variation around representative atomic centers. The results are shown in Fig.~3. The time-frequency spectra identify the dominant emission times of high-order harmonics, which correspond to the recollision of laser-driven electron wave packets with the molecular core.

For C$_{10}$H$_2$, pronounced harmonic emission occurs during the main recollision time window. Around these emission times, strong charge oscillations and electron localization are observed near the terminal carbon atoms. This indicates that the terminal regions of the long carbon chain play an important role in the HHG process. In contrast, for the short molecule C$_2$H$_2$, the charge redistribution is spatially more confined, and the electron acceleration distance is much shorter. As a result, the recollision probability and harmonic yield are both reduced.

These observations suggest that the enhanced HHG from C$_{10}$H$_2$ is closely related to its extended molecular backbone. The longer chain provides a larger spatial region for electron acceleration along the laser polarization direction and introduces additional atomic centers for possible recombination or scattering. This multi-center recollision process increases the efficiency of harmonic emission and contributes to the observed enhancement in the plateau region.

\begin{figure*}[t]
\centering
\includegraphics[width=1.0\textwidth]{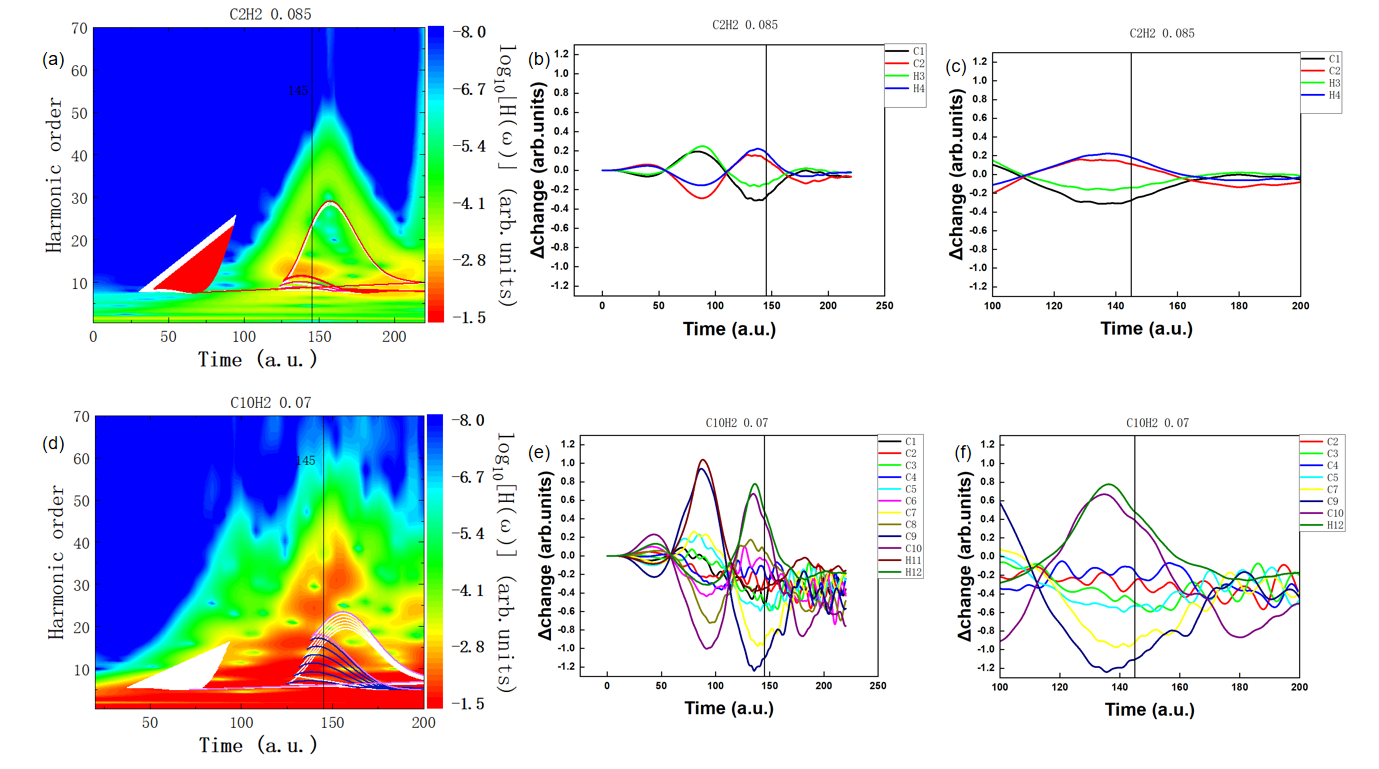}
\caption{
Time-frequency analysis and charge-density evolution for C$_2$H$_2$ and C$_{10}$H$_2$. 
(\textbf{a},\textbf{b}) HHG spectra and directional components. 
(\textbf{c},\textbf{d}) Time-frequency distributions of the HHG spectra, where the dashed lines mark the main harmonic emission times. 
(\textbf{e},\textbf{f}) Time-dependent charge variation around representative atomic centers. 
The stronger charge oscillations near the terminal carbon atoms of C$_{10}$H$_2$ indicate enhanced electron localization and multi-center recollision in the longer molecule.
}
\label{fig:3}
\end{figure*}


To examine whether the enhancement of HHG in longer carbon-chain molecules is robust with respect to laser intensity, we further calculate the HHG spectra of C$_2$H$_2$ and C$_{10}$H$_2$ at different field strengths. For each case, the laser amplitudes are selected to ensure the same Keldysh parameter for the two molecules. This allows a direct comparison of the harmonic response under comparable ionization regimes.

Figure~4 presents the HHG spectra obtained under different laser intensities. At relatively weak fields, the difference between C$_2$H$_2$ and C$_{10}$H$_2$ is modest, indicating that the molecular-size effect is not yet fully developed. As the laser intensity increases, the harmonic emission from C$_{10}$H$_2$ becomes increasingly stronger than that from C$_2$H$_2$, especially in the plateau and high-energy regions. This trend demonstrates that the size-dependent enhancement is more pronounced in stronger laser fields.

The enhanced emission from C$_{10}$H$_2$ can be attributed to its extended molecular geometry. A longer carbon chain provides a larger spatial region for electron acceleration and introduces more atomic centers for possible recollision. Consequently, the ionized electron wave packet can accumulate higher kinetic energy and has a greater probability of recombining or scattering from different nuclear centers. These effects jointly enhance the harmonic yield and modify the cutoff behavior.

\begin{figure*}[t]
\centering
\includegraphics[width=0.9667\textwidth]{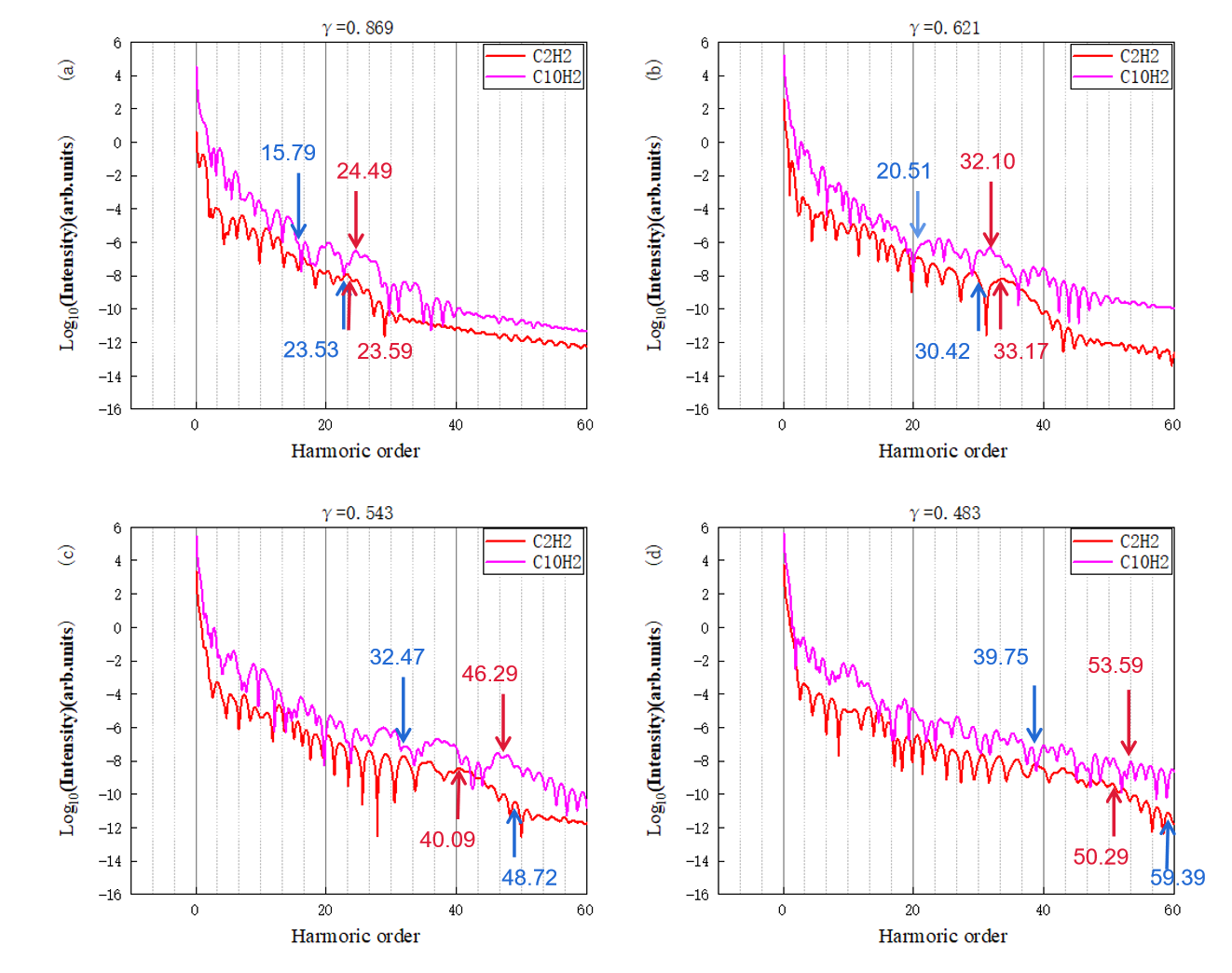}
\caption{
HHG spectra of C$_2$H$_2$ and C$_{10}$H$_2$ under linearly polarized laser fields with different intensities. For each panel, the laser amplitudes are chosen to maintain the same Keldysh parameter for the two molecules. The results show that the harmonic yield of C$_{10}$H$_2$ becomes increasingly stronger than that of C$_2$H$_2$ as the field strength increases.
}
\label{fig:4}
\end{figure*}


To further verify the size-dependent trend, we also calculate the HHG spectrum of the intermediate-chain molecule C$_6$H$_2$. Its molecular length lies between those of C$_2$H$_2$ and C$_{10}$H$_2$, making it suitable for examining whether the harmonic yield varies systematically with molecular size.
Figure~5(a) compares the HHG spectra of C$_2$H$_2$, C$_6$H$_2$, and C$_{10}$H$_2$ under the same Keldysh parameter. The harmonic yield of C$_6$H$_2$ lies between those of C$_2$H$_2$ and C$_{10}$H$_2$ over a broad spectral range, particularly in the plateau and high-energy regions. This result confirms that the enhancement of harmonic emission is strongly correlated with molecular chain length.
Further analysis of the time-frequency distribution and charge-density evolution of C$_6$H$_2$ provides additional evidence for this size-dependent mechanism. The dominant harmonic emission occurs around the recollision time window, accompanied by charge oscillations near the terminal carbon atoms. Compared with C$_2$H$_2$, the longer backbone of C$_6$H$_2$ provides a larger effective interaction length, while its response remains weaker than that of C$_{10}$H$_2$. Therefore, the three-molecule comparison establishes a clear relationship between molecular size and HHG efficiency.

To investigate the universality of enhanced harmonic emission in C\textsubscript{10}H\textsubscript{2} due to its larger molecular volume, we systematically varied the laser field intensity and calculated its high-harmonic spectra at four different strengths ranging from 0.05~a.u. to 0.09~a.u., as shown in Fig.~4a--d. For comparison, the harmonic spectra of C\textsubscript{2}H\textsubscript{2} were also computed. To preclude any influence from differences in ionization potential, all comparisons were conducted under conditions of an equal Keldysh parameter $\gamma$. The selected field strengths for C\textsubscript{10}H\textsubscript{2} correspond to $\gamma$ values of 0.869, 0.724, 0.621, 0.543, and 0.483. Using the Keldysh parameter formula, the equivalent field strengths required to maintain these $\gamma$ values for C\textsubscript{2}H\textsubscript{2} are 0.061~a.u., 0.073~a.u., 0.085~a.u., 0.098~a.u., and 0.110~a.u., respectively.

As shown in Fig.~4a--d, the harmonic cutoff position of C\textsubscript{2}H\textsubscript{2} consistently exceeds that of C\textsubscript{10}H\textsubscript{2} across all panels, a result attributed to their differing ionization potentials. However, the primary focus of this analysis is the contrast in harmonic intensity between the two species. The data from all four figures indicate that the harmonic intensity of C\textsubscript{10}H\textsubscript{2} is significantly higher than that of C\textsubscript{2}H\textsubscript{2}. Notably, as illustrated in Fig.~4a, under relatively low field strengths, the harmonic spectra of C\textsubscript{10}H\textsubscript{2} and C\textsubscript{2}H\textsubscript{2} exhibit almost no discernible difference.

These results demonstrate that the effective interaction length in linear carbon-chain molecules can be associated with the length of the carbon backbone. As the molecular chain length increases, the electron acceleration distance and multi-center recollision probability are enhanced, leading to stronger harmonic emission and a tendency toward cutoff extension.

\begin{figure*}[t]
\centering
\includegraphics[width=0.9333\textwidth]{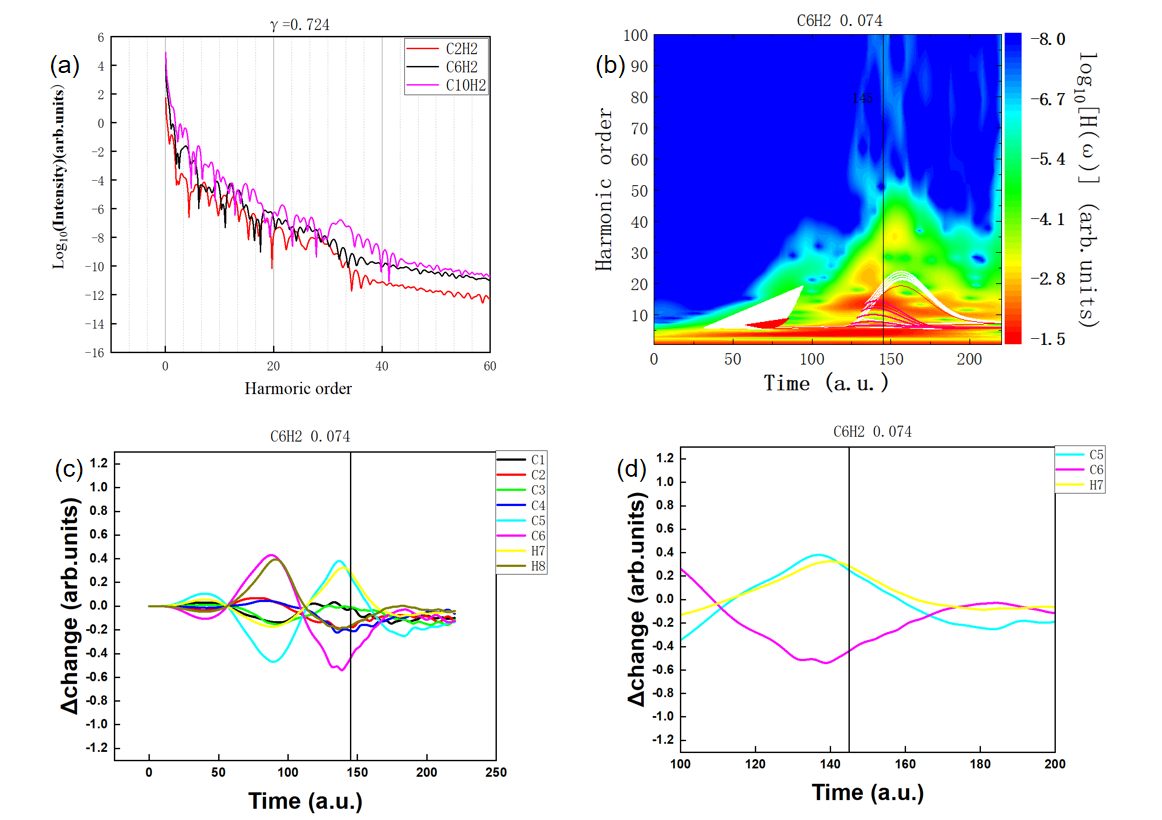}
\caption{
(\textbf{a}) Comparison of HHG spectra from C$_2$H$_2$, C$_6$H$_2$, and C$_{10}$H$_2$ under the same Keldysh parameter. 
(\textbf{b}) Time-frequency analysis of HHG from C$_6$H$_2$, where the dashed line marks the main harmonic emission time. 
(\textbf{c}) Time-dependent charge variation around representative atomic centers in C$_6$H$_2$ near the main emission time. 
The results show that the harmonic yield increases with molecular chain length and that terminal carbon atoms contribute strongly to the enhanced emission.
}
\label{fig:5}
\end{figure*}

\section{Theory and Method}\label{sec3}

The TDDFT method has proven to be a powerful tool for investigating molecular high-order harmonic generation in intense laser fields \cite{43,44,45}. Notably, its application to Xylene molecules \cite{46,47,48} and complex solid-state systems \cite{49,50} has demonstrated remarkable consistency with experimental observations.  

Based on the Runge--Gross theorem \cite{51}, the time-dependent external potential 
uniquely determines the time-dependent single-electron density for a many-body system 
evolving from a specified initial state. Under the length gauge and dipole approximation, 
the electron dynamics in molecular systems subjected to linearly polarized laser pulses are 
governed by the time-dependent Kohn--Sham (KS) equations for orbitals $\psi_i(\mathbf{r}, t)$, expressed as follows (atomic units are used throughout):  
\begin{equation}\resizebox{\dimexpr\columnwidth-2em\relax}{!}{$\displaystyle
\begin{aligned}
i \frac{\partial}{\partial t} \psi_i(\mathbf{r}, t)=\left[-\frac{1}{2} \nabla^2+V_{\mathrm{KS}}[\rho](\mathbf{r}, t)\right] \psi_i(\mathbf{r}, t), \quad i=1, \cdots N .
\end{aligned}
$}\end{equation}

In this formulation, \( i \) denotes the orbital index, while \( N \) represents the total 
number of Kohn--Sham orbitals. The time-dependent electron density, \( \rho(\mathbf{r}, 
t) \), is expressed as:  
\begin{equation}
\begin{aligned}
\rho(\mathbf{r}, t) = 2 \sum_{j=1}^N \left| \psi_j(\mathbf{r}, t) \right|^2.
\end{aligned}
\end{equation}

The Kohn--Sham potential \( V_{\mathrm{KS}}[\rho](\mathbf{r}, t) \), a functional of the 
electron density \( \rho(\mathbf{r}, t) \), is given by the following:
\begin{equation}\resizebox{\dimexpr\columnwidth-2em\relax}{!}{$\displaystyle
\begin{aligned}
V_{\mathrm{KS}}[\rho](\mathbf{r}, t)=V_{\mathrm{xc}}[\rho](\mathbf{r}, t)+V_{\mathrm{H}}[\rho](\mathbf{r}, t)+V_{\mathrm{ne}}(\mathbf{r})+V_{\text {laser }}(\mathbf{r}, t)
\end{aligned}
$}\end{equation}

The first term, \( V_{\mathrm{xc}}[\rho](\mathbf{r}, t) \), represents the 
exchange-correlation potential, which accounts for nonperturbative many-body effects and 
is modeled using the generalized gradient approximation (GGA) in the 
Perdew--Burke--Ernzerhof (PBE) form \cite{52}. The second term, \( 
V_{\mathrm{H}}[\rho](\mathbf{r}, t) \), corresponds to the Hartree potential, while the 
third term, \( V_{\mathrm{ne}}(\mathbf{r}) \), describes the electron-ion interactions, 
modeled using norm-conserving Troullier--Martins pseudopotentials \cite{53} in the 
Kleinman--Bylander parametrization \cite{54}. Lastly, \( V_{\mathrm{laser}}(\mathbf{r}, 
t) \) represents the external potential from the interaction between the molecule and the 
laser field, expressed as \( V_{\text{laser}}(\mathbf{r}, t) = \mathbf{r} \cdot \mathbf{E}(t) 
\), where \( \mathbf{E}(t) \) is the linearly polarized electric field of the laser.
\begin{equation}
\begin{aligned}
\mathbf{E}(t)=E f(t) \cos \left(\omega_0 t\right) \hat{\mathbf{e}}_{\mathrm{y}}
\end{aligned}
\end{equation}

The linearly polarized electric field has an amplitude of \( E = 0.06 \, \text{a.u.} \) and a wavelength of 800 nm. Its envelope is trapezoidal, with both the rising and falling edges lasting one optical cycle, and the plateau lasting four optical cycles.

The time-dependent Kohn--Sham orbital wavefunctions were propagated in real time on a 
real-space grid. The propagation method preserves time-reversal symmetry, and we 
employ the approximated enforced time-reversal symmetry (AETRS) \cite{55} with a time 
step of 0.08 a.u.

To mitigate unphysical effects arising from the reflection of the electron wave packet at the boundaries, a complex absorption potential (CAP) \cite{56} is used:
\begin{equation}\resizebox{\dimexpr\columnwidth-2em\relax}{!}{$\displaystyle
\begin{aligned}
\begin{aligned}
V_{\text {absorb }}(\mathrm{r})=\left\{\begin{array}{cl}
0, & 0<\mathrm{r}<\mathrm{r}_{\max } \\
i \eta \sin ^2\left[\frac{\left(\mathrm{r}-\mathrm{r}_{\max }\right) \pi}{2 L}\right], & \mathrm{r}_{\max }<\mathrm{r}<\mathrm{r}_{\max }+\mathrm{L}
\end{array}\right.
\end{aligned}
\end{aligned}
$}\end{equation}

Here, \( L = 10 \) a.u. and \( \eta = -0.8 \) a.u. represent the width and height of the absorbing potential, respectively.

The harmonic spectrum is derived from the time-dependent dipole acceleration \linebreak  \( \mathbf{a}(t) \) \cite{56}:
\begin{equation}
\begin{aligned}
H(\omega)=\left|\int\mathbf{a}(t) e^{-i \omega t} d t\right|^2
\end{aligned}
\end{equation}
where $\mathbf{a}(t)=-\int \rho(\mathbf{r}, t) \nabla H_{\mathrm{KS}} d^3 \mathbf{r}$,  $\quad H_{\mathrm{KS}}=-\frac{1}{2} \nabla^2+V_{\mathrm{KS}}[\rho](\mathbf{r}, t)$ . 

\section{Conclusions}\label{sec4}

In this work, we systematically investigated the influence of molecular chain length on HHG from linear carbon-chain molecules C$_2$H$_2$, C$_6$H$_2$, and C$_{10}$H$_2$ driven by an 800 nm linearly polarized laser field. By adjusting the laser intensity to keep the Keldysh parameter identical for different molecules, the influence of ionization-potential differences was minimized, allowing the role of molecular size to be isolated.

The TDDFT simulations show that the HHG yield increases significantly with molecular length. Compared with C$_2$H$_2$, the longer molecule C$_{10}$H$_2$ exhibits much stronger harmonic emission in the plateau region and a tendency toward cutoff extension. The intermediate molecule C$_6$H$_2$ shows a harmonic response between those of C$_2$H$_2$ and C$_{10}$H$_2$, further confirming the size-dependent enhancement of HHG in linear carbon-chain molecules.

Time-frequency analysis and charge-density evolution reveal that the enhanced HHG in longer chains is closely associated with electron localization and strong charge oscillations near the terminal carbon atoms. The extended molecular backbone provides a longer electron acceleration distance and increases the probability of multi-center recollision, thereby improving harmonic emission efficiency. These results demonstrate that molecular-size engineering is an effective strategy for controlling HHG in extended molecular systems and may provide useful guidance for developing intense attosecond light sources.





\authorcontributions{Conceptualization, M.H.; methodology, H.W.; software, H.W.; validation, N.X.; formal analysis, N.X.; investigation, D.Y.; resources, S.Z.; data curation, S.Z.; writing---original draft preparation, S.Z.; writing---review and editing, N.X.; supervision, M.H. All authors have read and agreed to the published version of the manuscript.}

\funding{This research was supported by the National Natural Science Foundation of China (NSFC) under Grant No. 12204214.}

\institutionalreview{Not applicable.}

\informedconsent{Not applicable.}

\dataavailability{The data presented in this study are available on request from the corresponding author.} 

\conflictsofinterest{The authors declare no conflicts of~interest.}




\end{document}